\documentclass[10pt]{elsart}
\usepackage{epsfig}
\newcommand{\bb}{b \bar b}
\newcommand{\bbb}{l \nu b b \bar b}
\newcommand{\bbbj}{l \nu b b \bar b j}
\newcommand{\etmiss}{E_T\!\!\!\!\!\!\!\!\! \not \,\,\,\,\,\,\,}
\newcommand{\ptmiss}{p_T\!\!\!\!\!\!\!\! \not \,\,\,\,\,\,\,}
\newcommand{\mtrec}{m_t^\mathrm{rec}}
\newcommand{\mtbrec}{m_{\bar t}^\mathrm{rec}}
\newcommand{\mhrec}{M_H^\mathrm{rec}}
\newcommand{\mwrec}{M_W^\mathrm{rec}}

\newcommand{\fbin}{fb$^{-1}$}

\begin{document}
\vspace*{-2cm}
\noindent
\hspace*{12.5cm}
UG--FT--113/00 \\
\hspace*{12.5cm}
FIST/4-2000/CFIF \\
\hspace*{12.5cm}
hep-ph/0004190 \\
\hspace*{12.5cm}
April 2000 \\
\begin{frontmatter}
\title{Probing top flavour-changing neutral scalar couplings at the CERN LHC}
\author{J. A. Aguilar-Saavedra} 
\address{Departamento de F\'{\i}sica Te\'{o}rica y del Cosmos \\
Universidad de Granada \\
E-18071 Granada, Spain}
\author{G. C. Branco}
\address{Departamento de Fisica and CFIF \\
Instituto Superior Tecnico \\
1096 Lisboa, Portugal}
\date{\today}
\begin{abstract}
Top decays into a light Higgs boson and an up or charm quark can reach
detectable levels in Standard Model extensions with two Higgs doublets or
with new exotic quarks, and in the Minimal Supersymmetric
Standard Model. Using both a standard and a neural network analysis we show
that the CERN Large Hadron Collider will give $3 \,\sigma$ evidence of decays
with $\mathrm{Br}(t \to Hc) \geq 6.5 \times 10^{-5}$ or set a limit
$\mathrm{Br}(t \to Hc) \leq 4.5 \times 10^{-5}$ with a 95\% confidence level if
these decays are not observed. We also consider limits obtained from
single top production associated with a neutral Higgs boson.

\vspace*{0.5cm} \noindent
PACS:  12.15.Mm; 12.60.Fr; 14.65.Ha; 14.80.Cp
% 12.15.Mm Neutral currents
% 12.60.-i Models beyond the standard model
% 12.60.Fr Extensions of electroweak Higgs sector
% 12.60.Jv Supersymmetric models
% 14.65.Ha Top quarks
% 14.80.Cp Non-standard-model Higgs bosons

\end{abstract}
\end{frontmatter}
%\twocolumn
\section{Introduction}
The search for flavour-changing neutral (FCN) current processes both at high and
low energies is one of the major tools to test the Standard Model (SM), with the
potential for either discovering or putting stringent limits on new physics. In
the SM, there are no gauge FCN couplings at tree level due to the GIM mechanism,
and scalar couplings are also automatically flavour-diagonal provided only one
Higgs doublet is introduced. However, there is no fundamental reason for having
only one Higgs doublet, and for two or more scalar doublets FCN couplings are
generated at tree level unless an {\em ad hoc\/} discrete symmetry is imposed
\cite{papiro1}.

There are stringent limits on the FCN couplings between light quarks arising
from the smallness of the $K^0 - \bar K^0$, $B^0 - \bar B^0$ and $D^0 - \bar
D^0$ mass differences \cite{papiro2}, as well as from the CP violation parameter
$\epsilon$ of the neutral kaon system. If there is no suppression of scalar FCN
couplings, the masses of the neutral Higgs bosons
have to be of the order of 1 TeV.
However, if FCN couplings between light quarks are suppressed, being for
instance
proportional to the quark masses \cite{papiro2b,papiro3} or to some combination
of Cabibbo-Kobayashi-Maskawa (CKM) mixing angles \cite{papiro3b,papiro4},
the lightest neutral Higgs may have a mass $M_H$ of around 100 GeV. In this
case, the effects of scalar FCN currents involving the top quark can be quite
sizeable. 

The interactions among the top, a light quark $q=u,c$
and a Higgs boson $H$ will be described by the effective Lagrangian
\begin{equation}
\mathcal{L} = \frac{g_W}{2 \sqrt 2} g_{tq} \bar q (c_v+c_a \gamma_5) t H ~~+~
\mathrm{h.c.}
\end{equation}
The axial and vector parts of the coupling are normalized to
$|c_v|^2+|c_a|^2 = 1$.
The natural size of $g_{tc}$ in a two Higgs doublet model (2HDM) is
$g_{tc} \sim 0.20$, leading to large FCN current effects observable at the CERN
Large Hadron Collider (LHC). The natural size of $g_{tu}$ ranges from 0.012 to
0.025, depending on the model considered. This is too small to
be observed, but if this coupling is enhanced by a factor
$\sim 2$, it could also be detected.
Models with new heavy exotic quarks can also have large tree level $Htq$
couplings, with the particularity that they are proportional to the FCN $Ztq$
couplings \cite{papiro4b}. Present limits on the latter \cite{papiro4c}
allow for $g_{tu} \sim 0.31$ or $g_{tc} \sim 0.35$, both observable at LHC.

The large top quark mass opens the possibility of having relatively large
effective vertices $Htq$ induced at one loop by new physics. In the
Minimal Supersymmetric Standard Model (MSSM) there are large regions of the
parameter space where one can have $g_{tc} \sim 0.04$ \cite{papiro4d,papiro5}
\footnote{The value of $g_{tu}$ is much smaller in principle.}.
In contrast, in the SM the effective $Htc$ vertex is very small,
$g_{tc} \sim 10^{-6}$ due to a strong GIM cancellation \cite{papiro6}.
The large top quark mass also suggests that the top quark could play a
fundamental r\^ole in the electroweak symmetry breaking mechanism
\cite{papiro8,papiro9} and in this case top FCN scalar couplings would be large,
$g_{tc} \sim 0.3-0.4$ \cite{papiro9b} or $g_{tc} \sim 0.5-2.9$ \cite{papiro9c},
depending on the model considered.
Hence the importance of measuring its couplings, in
particular to the Higgs boson \cite{papiro7}.

Present experimental bounds on top FCN scalar (and gauge) couplings are
very weak. Some processes have been proposed to measure the $Htq$ vertices, 
for instance $t \bar c \nu_e \bar \nu_e$ and $t \bar c e^+ e^-$ production at a
linear collider with a center of mass (CM) energy of $1-2$ TeV
\cite{papiro10} and $t \bar c$ production at a muon collider \cite{papiro10b}
or at hadron colliders via gluon fusion $gg \to H \to t \bar c$
\cite{papiro9b}.
These processes can probe the FCN couplings of
the heavier mass eigenstates, but do not provide useful limits for the light
scalars. In this Letter we will show that top decays $t \to Hq$ at LHC provide
the best limits on the couplings of a light scalar of a mass slightly above
100 GeV. The importance of this mass range stems from the fact that
global fits to
electroweak observables in the SM \cite{papiro11} seem to prefer $M_H \sim 100$
GeV, as well as the MSSM prediction that the mass of the lightest
scalar has to be less than 130 GeV \cite{papiro12}. On the other hand,
direct searches at the CERN $e^+ e^-$ collider LEP
at CM energies up to 202 GeV imply the bound $M_H > 107.9$ GeV for the
mass of a SM Higgs, with a 95\% confidence level (CL) \cite{papiro13}. For the
lightest Higgs of the MSSM the bounds depend on the choice of parameters and
typically are reduced to $M_H > 88.3$ GeV.

In the following we will perform a detailed analysis of the sensitivity of the
LHC to FCN scalar couplings in top decays $t \to Hq$ \cite{papiro14a}.
We will also study $Ht$
production, which also serves to constrain the $Htq$ vertex
\cite{papiro14}. However, the bounds obtained from this process are much less
restrictive. Finally, we will comment on the potential of the LHC to
discover these signals at the rates predicted by some models.

\section{Limits on FCN couplings from top decays}

To obtain constraints on the coupling $g_{tq}$ we will first consider $t
\bar t$ production, with the top quark decaying to $W^+b \to l \nu b$, 
$l=e,\mu$ and the antitop decaying to $H \bar q$.
With a good $\tau$ identification the decays $W^+ \to \tau^+ \nu$ can be
included, improving the statistical significance of the signal.

The decay mode of the Higgs boson depends strongly on its mass.
For our evaluation, we take $M_H = 120$ GeV, well
above the LEP limits, and decaying into $\bb$ pairs.
In principle, there is no reason to assume that the $Hbb$ coupling is
suppressed. In a general 2HDM,
both doublets couple to the $b$ quark and, unless
some fine-tuned cancellation occurs, the coupling of the physical light Higgs to
the $b$ quark is proportional to the $b$ mass, which we take $m_b=3.1$ GeV
at the scale $\mu = M_H$. To be conservative, we assume that
the Higgs couplings to the gauge bosons are not suppressed and are the same as
in the SM. Thus for $M_H=120$ GeV $H$ decays predominantly into $\bb$ pairs,
with a branching ratio of $0.7$.
For $M_H > 130$ GeV, the $W^+ W^-$ decay mode, with
one of the $W$'s off-shell, begins to dominate
\footnote{Of course, if the
couplings to the gauge bosons are suppressed, for instance if $H$ is a
pseudoscalar, the signal will get an
enhancement factor of roughly $9/7$ and the $\bb$ decay mode will dominate
also for larger values of $M_H$.}.
The signal is then $\bbbj$, with three $b$ quarks in the final
state. Clearly, $b$ tagging is crucial to separate the signal
from the backgrounds. We assume a $b$ tagging efficiency of 50\%, and a mistag
rate of 1\%, similar to the values recently achieved at SLD
\cite{papiro21}. 

We evaluate the signal using the full
$gg, q\bar q \to t \bar t \to W^+b H \bar q \to  l \nu b \bb q$
matrix elements with all intermediate particles off-shell.
We calculate the matrix elements using HELAS \cite{papiro15} code generated by
MadGraph \cite{papiro16} and modified to include the decay chain. For
definiteness we assume $c_v=1$, $c_a = 0$, but we check that the difference 
when using other combinations is below $1\%$.
We
normalize the cross section with a $K$ factor of 2.0 \cite{papiro18}.

The main background to this signal comes from $t \bar t$ production with
standard top
and antitop decays, $t \to W^+b \to l \nu b $,
$\bar t \to W^- \bar b \to jj \bar b$.
This process mimics the signal if one of the jets resulting from the $W^-$ decay
is mistagged as a $b$ jet.
The matrix elements for this process are calculated with
HELAS and MadGraph in the same way as the signal, taking again $K=2.0$.
Another
background to be considered is $W \bb jj$ production, for which we use VECBOS
\cite{papiro17} modified to include energy smearing and trigger and
kinematical cuts. For this process we take $K=1.1$ \cite{papiro19}.
$Wjjjj$ production is negligible after using $b$ tagging. For the phase space
integration we use MRST structure
functions set A \cite{papiro20} with $Q^2=\hat s$.

After generating signals and backgrounds, we simulate the detector resolution
effects with a Gaussian smearing of the energies of jets $j$ and charged leptons
$l$ \cite{papiro22},
\begin{equation}
\frac{\Delta E^{l}}{E^{l}} = \frac{10\%}{\sqrt{E^{l}}}
\oplus 0.3\% \,, ~~~~
\frac{\Delta E^{j}}{E^{j}} = \frac{50\%}{\sqrt{E^j}} \oplus 3\% \,,
\end{equation}
where the energies are in GeV and
the two terms are added in quadrature. We then apply
detector cuts on transverse momenta $p_T$, pseudorapidities $\eta$ and distances
in $(\eta,\phi)$ space $\Delta R$:
$$
p_T^l \geq 15 ~\mathrm{GeV} ~,~~ p_T^j \geq 20 ~\mathrm{GeV}
$$
\begin{equation}
|\eta^{l,j}|  \leq 2.5 ~,~~
\Delta R_{jj,lj}  \geq 0.4 \,.
\end{equation}
For the events to be triggered, we require both the signal and background
to fulfill at least one of the trigger conditions \cite{papiro23}. These
conditions are different in the low 10 \fbin\ (L) and high 100 \fbin\ (H)
luminosity Runs. For our
processes, the relevant conditions are
\begin{itemize}
\item one jet with $p_T \geq 180$ GeV (L) / 290 GeV (H),
\item three jets with $p_T \geq 75$ GeV (L) / 130 GeV (H),
\item one charged lepton with $p_T \geq 20$ GeV (L) / 30 GeV (H),
\item missing energy $\etmiss \geq 50$ GeV (L) / 100 GeV (H) and one jet with
$p_T \geq 50$ GeV (L) / 100 GeV (H).
\end{itemize}

\begin{figure}[tb]
\begin{center}
\mbox{\epsfig{file=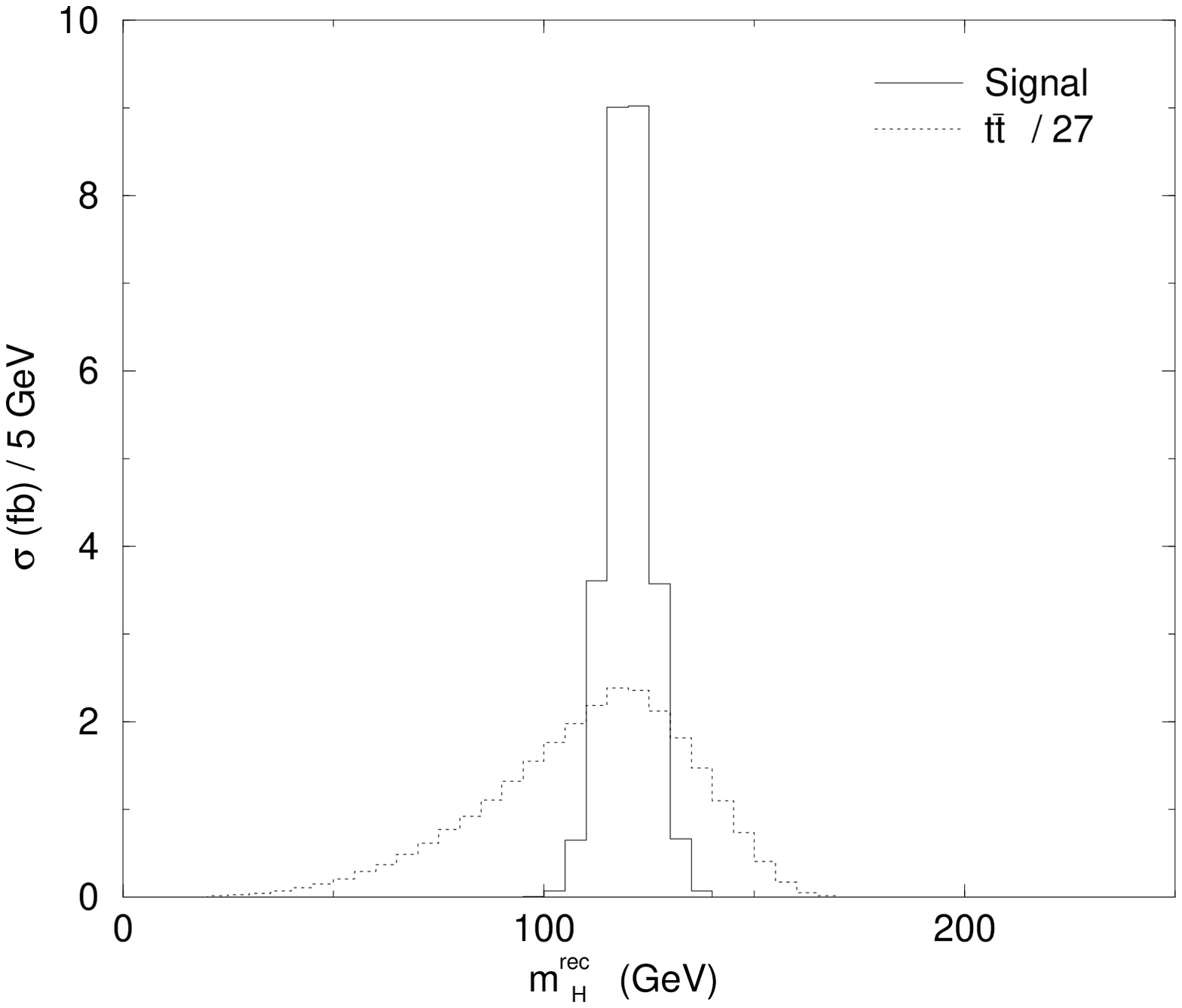,width=10cm}}
\end{center}
\caption{Reconstructed Higgs mass $\mhrec$ distribution before kinematical cuts
for the signal and $t \bar t$ background in LHC Run L.
We use  $g_{tq} = 0.2$.
\label{fig:1} }
\end{figure}

\begin{figure}[tb]
\begin{center}
\mbox{\epsfig{file=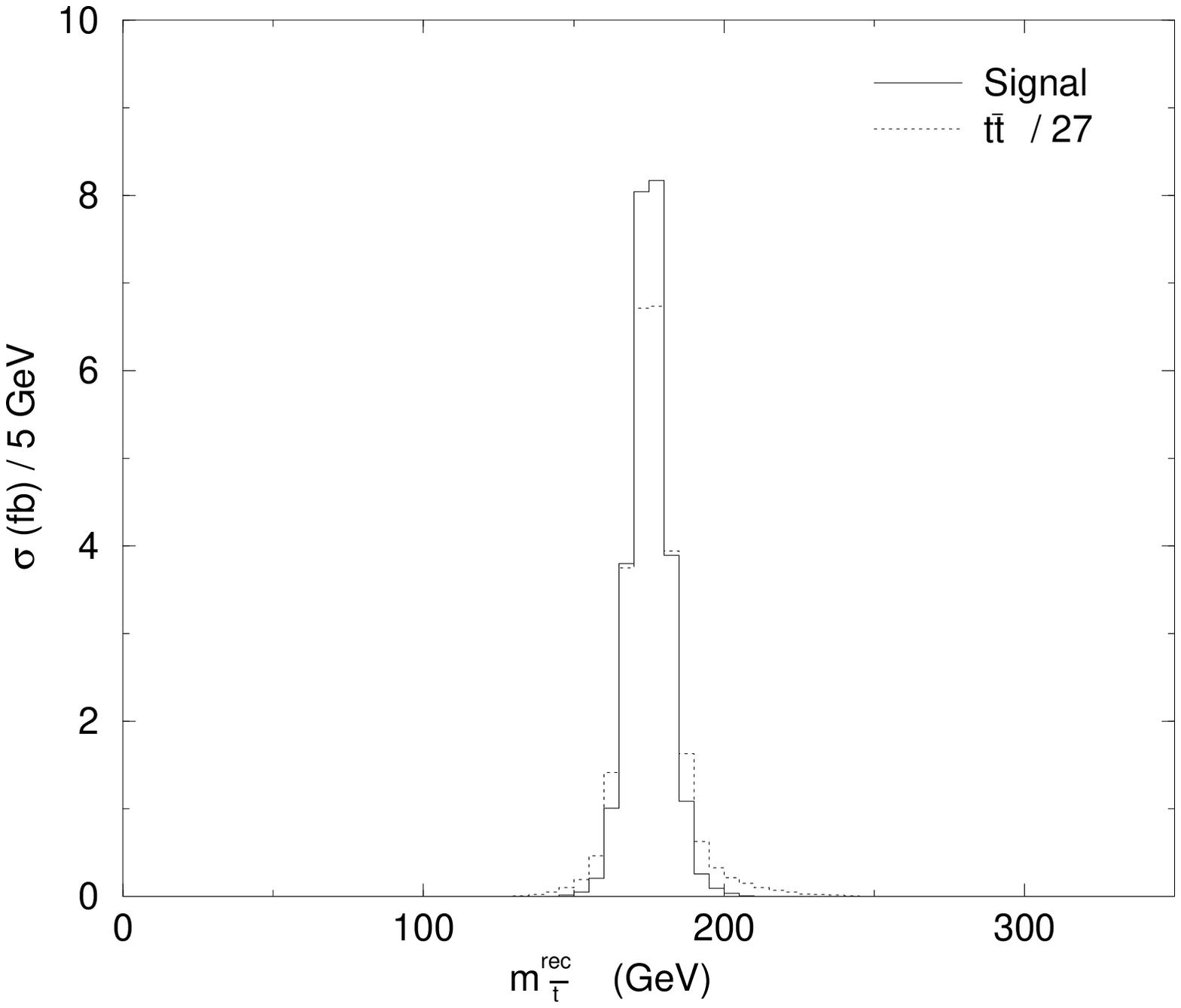,width=10cm}}
\end{center}
\caption{Reconstructed antitop mass $\mtbrec$ distribution before kinematical
cuts for the signal and $t \bar t$ background in LHC Run L.
We use  $g_{tq} = 0.2$.
\label{fig:2} }
\end{figure}

We reconstruct the signal as follows. There are three tagged $b$ jets $b_{1-3}$
in the final state and a non-$b$ jet $j$. There are three possible pairs $b_i,
b_j$, with invariant masses $M_{b_i b_j}$, one of which results from
the decay of the Higgs boson and has an invariant mass close to $M_H$.
The invariant mass of this pair and the jet $j$, $M_{b_i b_j j}$, is also
close to the top mass. Hence we choose the pair $b_i,b_j$
which minimizes $(M_{b_i b_j}-M_H)+(M_{b_i b_j j}-m_t)$, defining the Higgs
reconstructed mass as $\mhrec = M_{b_i b_j}$ and the antitop reconstructed mass
as $\mtbrec = M_{b_i b_j j}$. In Figs~\ref{fig:1},~\ref{fig:2}
 we plot the kinematical
distributions of these variables for the signal and for the
$t \bar t$ background. The
$\mtbrec$ distribution is slightly broader for $t \bar t$ due to the signal
reconstruction method.

\begin{figure}[tb]
\begin{center}
\mbox{\epsfig{file=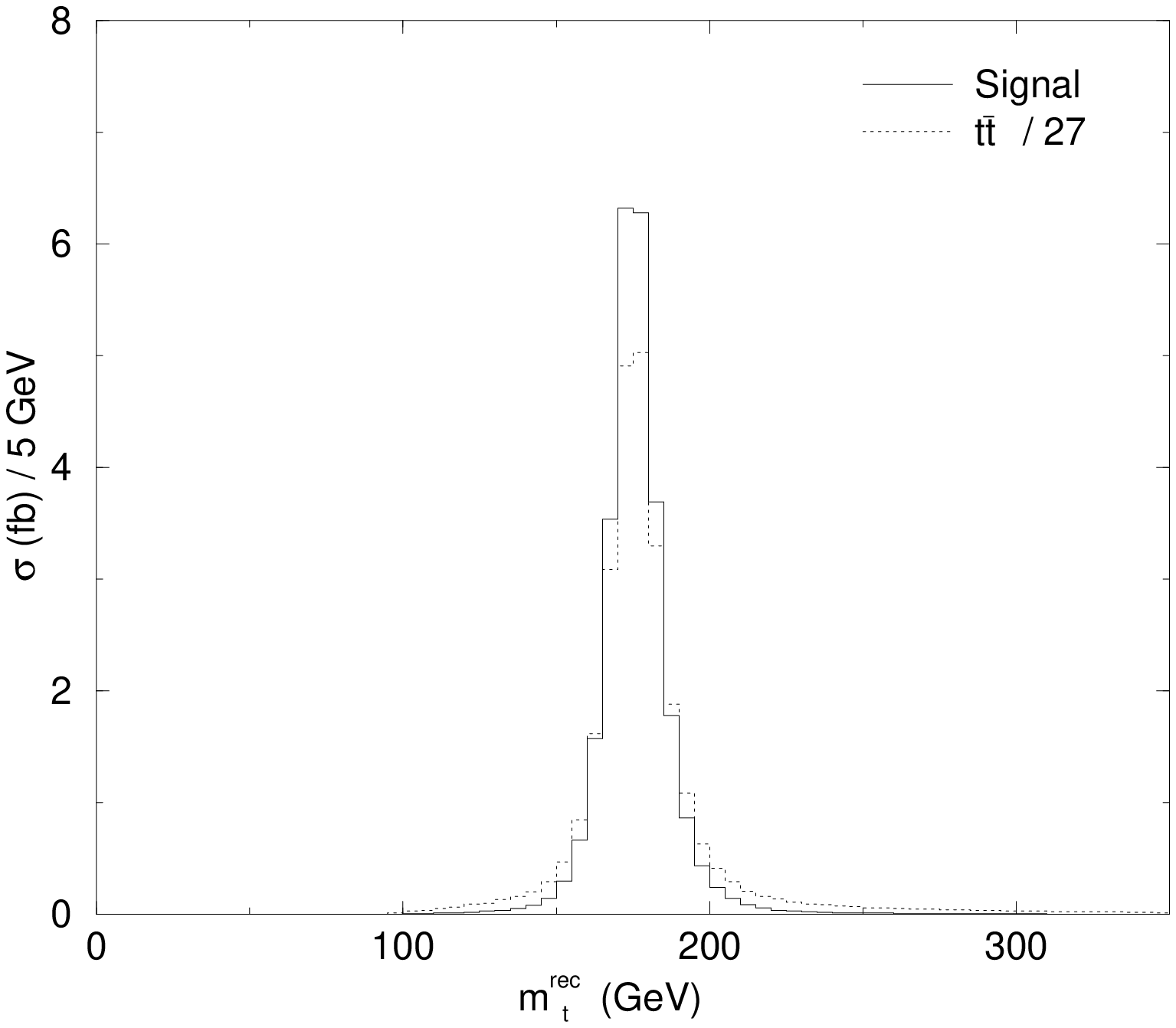,width=10cm}}
\end{center}
\caption{Reconstructed top mass $\mtrec$ distribution before kinematical
cuts for the signal and $t \bar t$ background in LHC Run L.
We use  $g_{tq} = 0.2$.
\label{fig:3} }
\end{figure}

The remaining jet $b_k$ and the charged lepton result
from the decay of the top quark. To reconstruct
its mass, we make the hypothesis that all missing energy comes from a
single neutrino with
$p^\nu=(E^\nu,\ptmiss,p_L^\nu)$, and $\ptmiss$ the missing transverse momentum.
Using $(p^l + p^\nu)^2 = M_W^2$ we find two
solutions for $p^\nu$, and we choose that one making the reconstructed top mass
$\mtrec \equiv \sqrt{(p^l +  p^\nu + p^{b_k})^2}$ closest to $m_t$.
In Fig.~\ref{fig:3} we plot the kinematical distribution of this variable for
the signal and $t \bar t$ background.

\begin{figure}[tb]
\begin{center}
\mbox{\epsfig{file=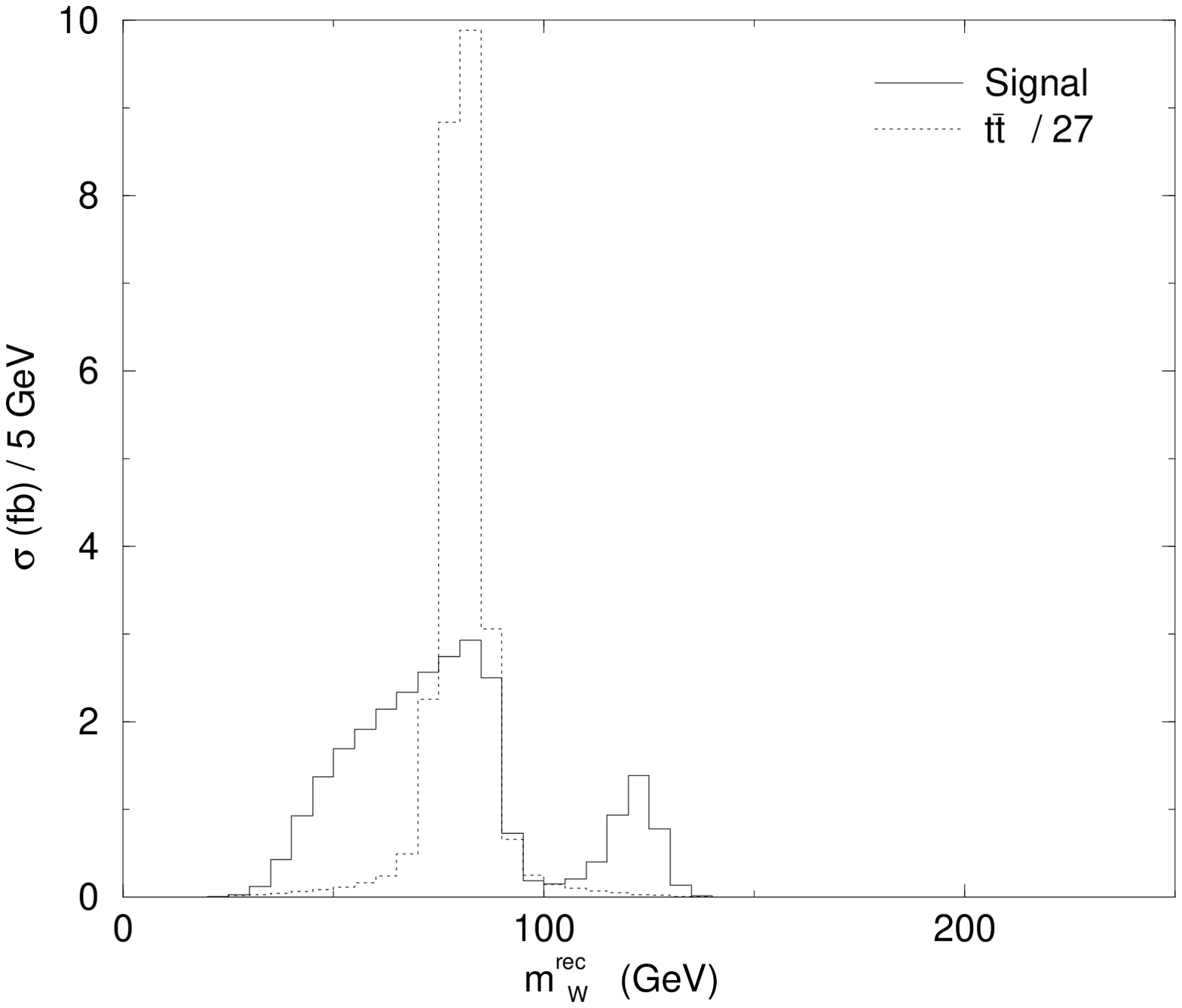,width=10cm}}
\end{center}
\caption{Reconstructed $W$ mass $\mwrec$ distribution before kinematical
cuts for the signal and $t \bar t$ background in LHC Run L.
We use  $g_{tq} = 0.2$.
\label{fig:4} }
\end{figure}

The reconstruction of the signal and the requirement $\mtrec,\mtbrec \sim m_t$,
$\mhrec \sim M_H$ are sufficient to eliminate the $W\bb jj$ background, but
hardly affect $t \bar t$, as can be seen in Figs.~\ref{fig:1}--\ref{fig:3}.
To improve the signal to background ratio we reject events when one pair of
jets, $b_i j$ or $b_j j$, seems to be the product of the
hadronic decay of a $W$ boson,
as happens for the $t \bar t$ background. We define the reconstructed $W$ mass
as the invariant mass $M_{b_i j}$, $M_{b_j j}$ closest to $M_W$ (see
Fig.~\ref{fig:4}) and impose a veto cut on events with $\mwrec \sim M_W$.

The complete set of kinematical cuts for both Runs is summarized in
Table~\ref{tab:1}. We also require a large transverse energy $H_T$.

\begin{table}[htb]
\begin{center}
\begin{tabular}{cc}
Variable & Cut \\
$\mhrec$ & 110--130 \\
$\mtbrec$ & 160--185 \\
$\mtrec$ & 160--190 \\
$\mwrec$ & $<65$ or $>110$ \\
$H_T$ & $>240$ \\
\end{tabular}
\caption{Standard kinematical cuts for the $\bbbj$ signal.
The masses and the energy are in GeV.
\label{tab:1}}
\end{center}
\end{table}

\begin{figure}[tb]
\begin{center}
\mbox{\epsfig{file=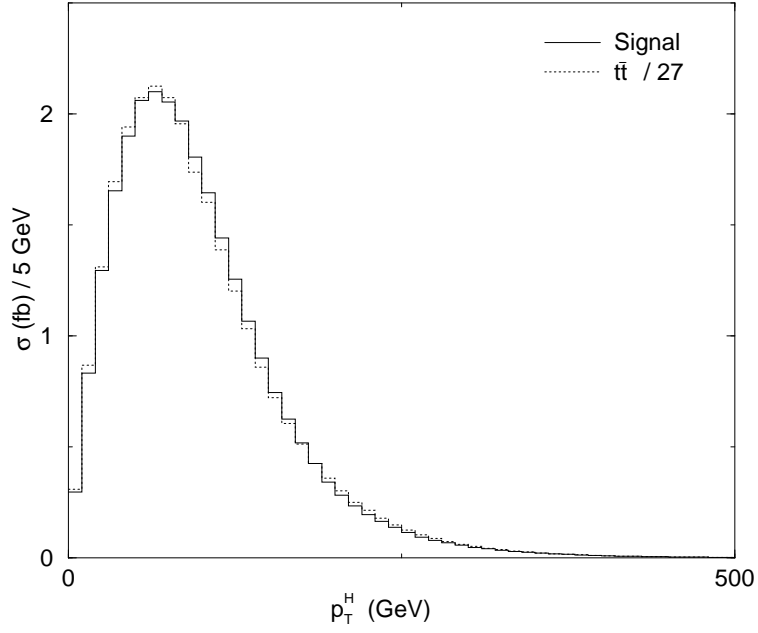,width=10cm}}
\end{center}
\caption{Higgs transverse momentum $p_T^H$ distribution before kinematical
cuts for the signal and $t \bar t$ background in LHC Run L.
We use  $g_{tq} = 0.2$.
\label{fig:5} }
\end{figure}

\begin{figure}[tb]
\begin{center}
\mbox{\epsfig{file=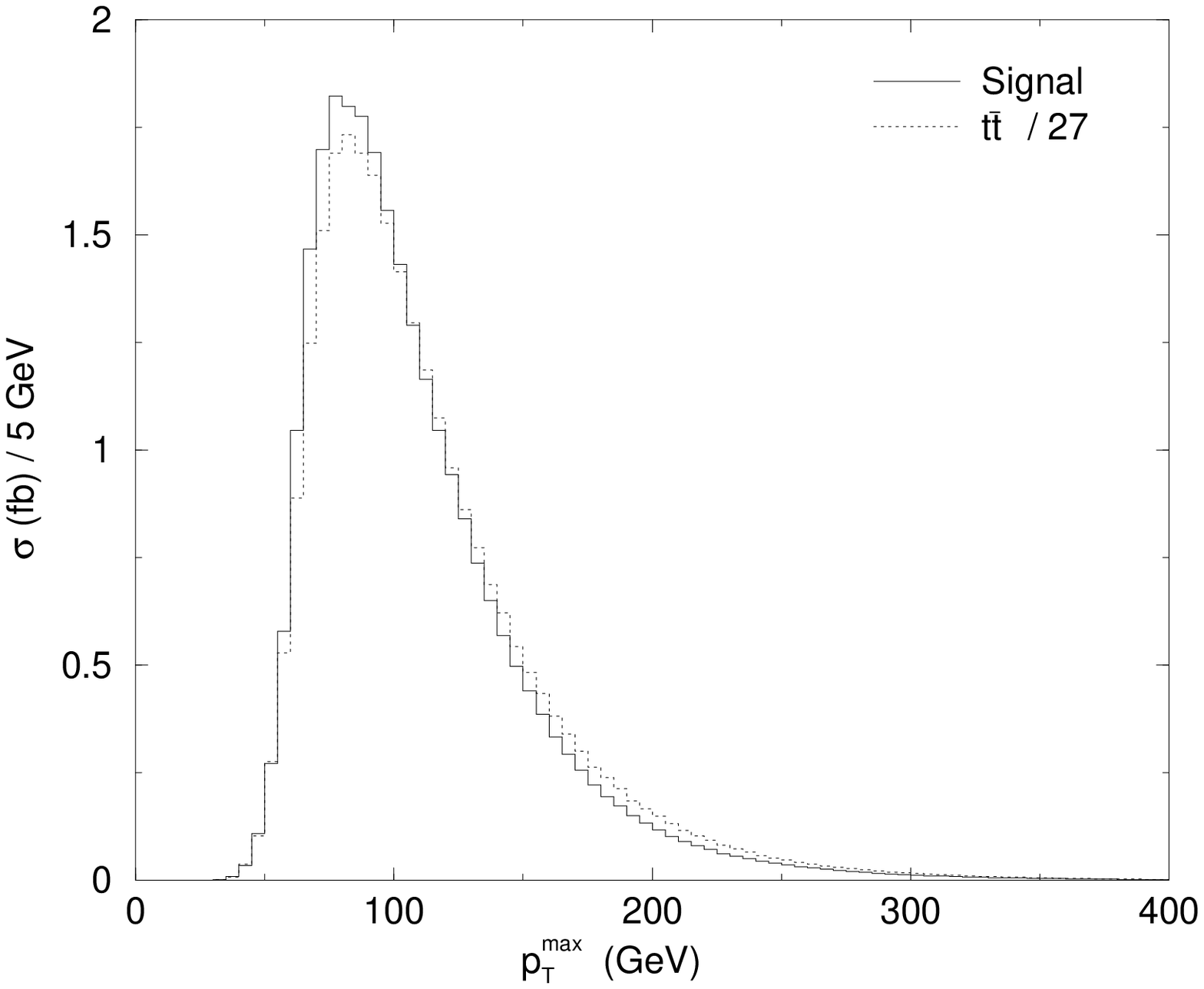,width=10cm}}
\end{center}
\caption{$p_T^\mathrm{max}$ distribution before kinematical
cuts for the signal and $t \bar t$ background in LHC Run L.
We use  $g_{tq} = 0.2$.
\label{fig:6} }
\end{figure}

\begin{figure}[tb]
\begin{center}
\mbox{\epsfig{file=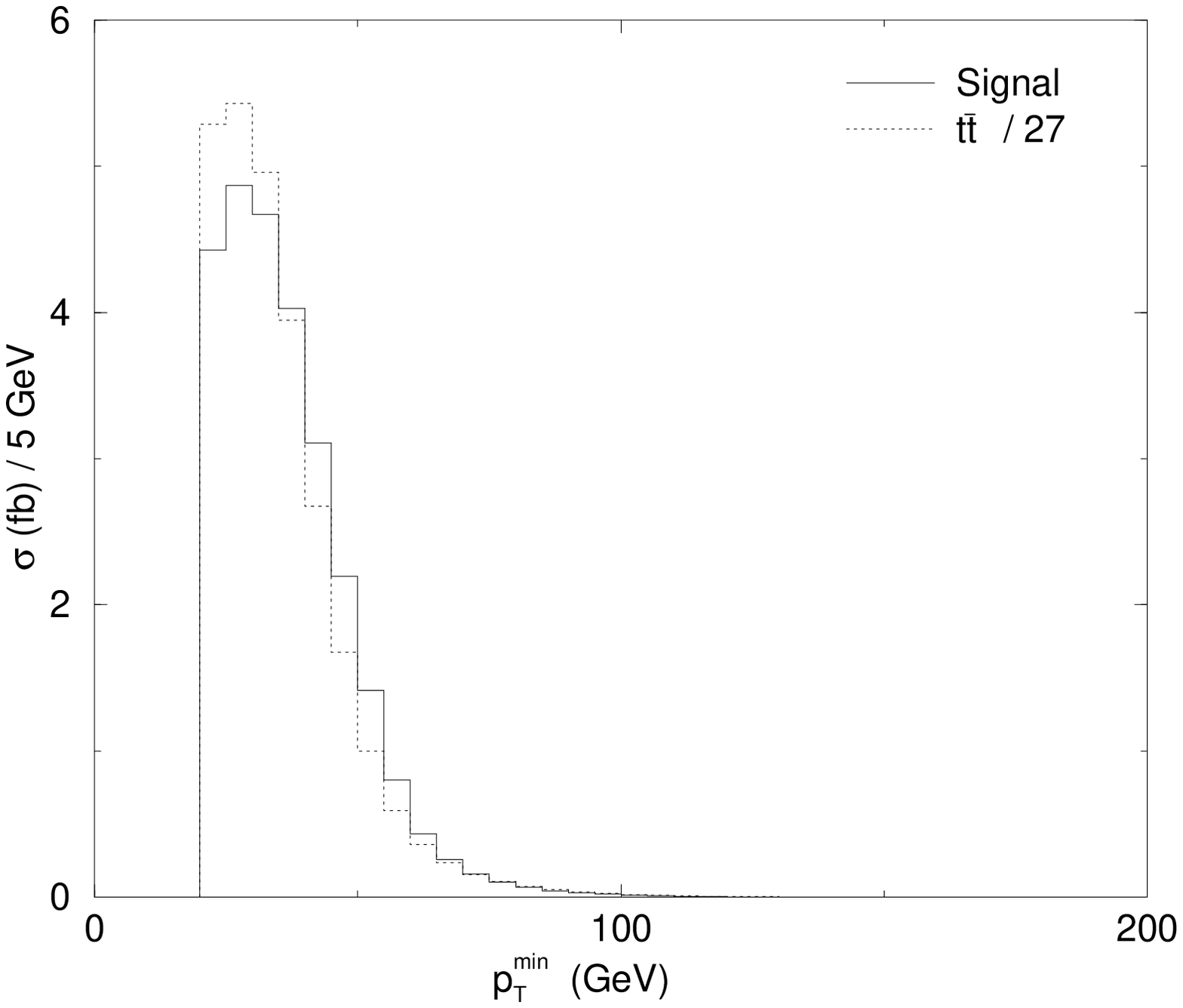,width=10cm}}
\end{center}
\caption{$p_T^\mathrm{min}$ distribution before kinematical
cuts for the signal and $t \bar t$ background in LHC Run L.
We use  $g_{tq} = 0.2$.
\label{fig:7} }
\end{figure}

\begin{figure}[tb]
\begin{center}
\mbox{\epsfig{file=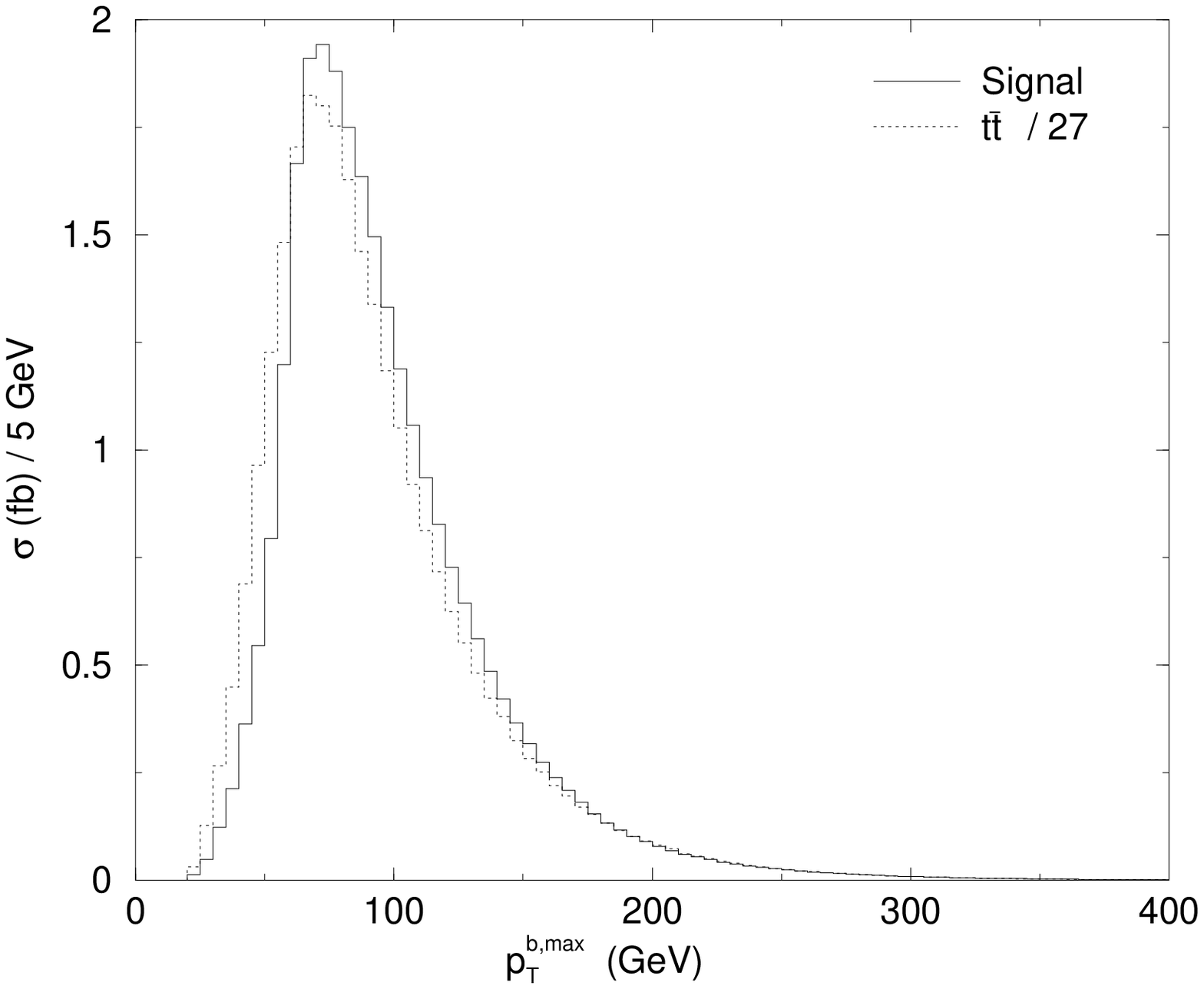,width=10cm}}
\end{center}
\caption{$p_T^{b,\mathrm{max}}$ distribution before kinematical
cuts for the signal and $t \bar t$ background in LHC Run L.
We use  $g_{tq} = 0.2$.
\label{fig:8} }
\end{figure}

Alternatively, we can use an artificial neural network (ANN) as a classifier to
distinguish between the signal and the $t \bar t$ background. In addition to
$\mhrec$, $\mtbrec$, $\mtrec$ and $\mwrec$, we consider $p_T^H$, the transverse
momentum of the reconstructed Higgs boson, $p_T^\mathrm{max}$ and
$p_T^\mathrm{min}$, the maximum and minimum transverse momentum of the jets, 
and $p_T^{b,\mathrm{max}}$, the maximum transverse momentum of the two $b$'s
which reconstruct the Higgs boson. The kinematical distributions of these
variables are shown in Figs.~\ref{fig:5}--\ref{fig:8}.

We observe that these variables are clearly not suitable to perform
kinematical cuts on them. However, their inclusion as inputs to an ANN greatly
improves its ability to classify signal and background events. We have not found
any improvement adding other
variables, and in some cases the results are worse.

To construct the ANN we use JETNET 3.5 \cite{papiro24}. We find convenient using
a three-layer topology of 8 input nodes, 12 hidden nodes and 1 output node.
However, we do not claim that either the choice of variables or the network
topology are the best ones. The performance of the ANN is not very sensitive to
the number of hidden nodes. The
input variables are normalized to lie approximately in the interval $[-1,1]$ to
reduce the training time. The network output $r$ is set to one for the signal
and zero for $t \bar t$.
We train the network using two sets with roughly 40000 signal and 40000
background events and the standard backpropagation algorithm. To test the
ability of the network to classify never seen data we use two other sets with
the same size. 

One frequent problem is overtraining. To avoid it, we keep training while the
network error evaluated on the test sets,
$E_t^{(N)}$, decreases for the successive training epochs $N$. When it begins to
increase, we stop training. To avoid fluctuations, we keep track of $E_t^{(N)}$
for previous epochs.
When we reach an epoch $N_0$ when $E_t^{(N_0-100)}$ is smaller
than all the following values, we stop training and start again until epoch
$N_0-100$.

We repeat the same procedure using four other different training and test sets
of the same size, and find that the results are fairly stable. The final
training is done including the 4 training and 4 test sets. At the end of the
training, most of the signal and background events in the test samples
concentrate in very narrow intervals $[0.99,1]$ and $[0,0.01]$, respectively.

%\begin{figure}[htb]
%\begin{center}
%\mbox{\epsfig{file=figs/RNET0.eps,width=10cm}}
%\end{center}
%\caption{Network output $r$ distribution for the test samples used in the
%training procedure (each set has $\sim 80000$ events). The $Y$ axis represents
%the number of events of these samples {\em without} including the cross
%section.
%\label{fig:9a} }
%\end{figure}

To classify signal and background we use two other signal and background sets
with never seen before data, 50 and 100 times larger, respectively, than the
ones used in training, to avoid possible statistical fluctuations.
%
% The distribution of the
% network output $r$ on these sets weighted with the cross section is shown in
% Fig.~\ref{fig:9}.
%
%\begin{figure}[tb]
%\begin{center}
%\mbox{\epsfig{file=figs/RNET.eps,width=10cm}}
%\end{center}
%\caption{Network output $r$ distribution for the signal and $t \bar t$
%background in LHC Run L. The $Y$ axis represents the cross section per bin.
%We use  $g_{tq} = 0.2$.
%\label{fig:9} }
%\end{figure}
To separate signal from background we simply require $r>0.998$.
 This maintains
29\% of the signal while it rejects 99.9\% of the $t \bar t$
background. It is not necessary to train another ANN to distinguish the
$W \bb jj$ background, the same network with $r>0.998$
 completely eliminates it. In Table~\ref{tab:2} we
collect the number of events without cuts, with the standard cuts in
Table~\ref{tab:1} and with the ANN cut $r>0.998$,
for LHC Runs L and H. We
normalize the signal to $g_{tq} = 0.2$.

\begin{table}[htb]
\begin{center}
\begin{tabular}{ccccccc}
 & \multicolumn{3}{c}{Run L} & \multicolumn{3}{c}{Run H} \\
& before & standard & ANN & before & standard & ANN \\[-0.4cm]
& cuts & cuts & cuts & cut & cuts & cut \\
Signal & 267 & 98.2 & 76.2 & 2150 & 797 & 614 \\
$t \bar t$ & 7186 & 33.2 & 10.0 & 58230 & 270 & 80 \\
$W \bb jj$ & 77 & 0.3 & 0.1 & 644 & 2.2 & 1.0 \\
\end{tabular} 
\caption {Number of $\bbbj$ events before and after
kinematical cuts for the signal and backgrounds.
We use $g_{tq}=0.2$.
\label{tab:2}}
\end{center}
\end{table}

Although for definiteness we train the ANN with a vector FCN coupling ($c_v=1$,
$c_a=0$), we check that the same ANN correctly recognizes the signal
events when we consider an axial, left- or right-handed FCN coupling. The
differences between the cross sections are in all cases below $1\%$, and
comparable to the statistical Monte Carlo uncertainty. The differences between
the cross sections after standard kinematical cuts are also smaller than $1\%$.

To derive upper bounds on the FCN couplings we follow the Feldman-Cousins
construction \cite{papiro25}. For the numerical evaluation of the confidence
intervals we use the PCI package \cite{papiro26}.
If there is no evidence of this process, {\em i. e.}, if the number of events
observed $n_0$ is equal to the expected background $n_b$, we obtain independent
limits on the couplings $g_{tu}$, $g_{tc}$. Using the standard cuts we obtain
$g_{tq} \leq 0.071$ at Run L and $g_{tq} \leq 0.041$ at Run H. Using the ANN
cut improves these limits, $g_{tq} \leq 0.064$ at Run L and
$g_{tq} \leq 0.035$ at Run H.
All bounds are calculated with a 95\% CL.
Although the improvement may not seem very significant,
in Section \ref{sec:4}
we will see that the luminosity required to see a positive signal is
reduced to one half with the help of the neural network.

It is worth to give here also the results for the Fermilab Tevatron. This
process at Run II with a CM 
energy of 2 TeV and a luminosity of 20 \fbin\ gives only $g_{tq} \leq 0.38$
(using standard kinematical cuts) due to the smaller statistics available.

\section{Limits on FCN couplings from $Ht$ production}

\begin{figure}[htb]
\begin{center}
\mbox{\epsfig{file=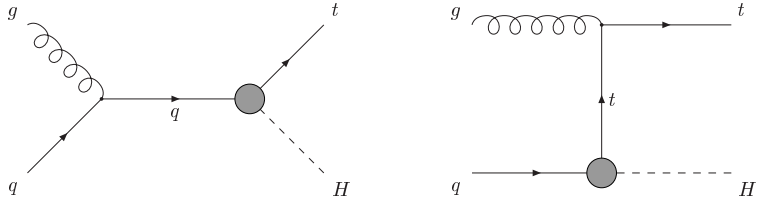,width=10cm}}
\end{center}
\caption{Feynman diagrams for $gq \to Ht$ via FCN couplings.
\label{fig:10}}
\end{figure}

The existence of top FCN Higgs couplings leads to $Ht$ production through the
diagrams in Fig.~\ref{fig:10}. This process has a negligible cross section in
the SM \cite{papiro29}.
As in the previous section, we consider only the
leptonic decays of the top quark, and the signal is $\bbb$. Note that
$Ht$ production is completely analogous to $Zt$
production via $Ztq$ anomalous couplings with $Z \to \bb$ \cite{papiro27},
and has the same large backgrounds as this decay mode.
The most dangerous is $t \bar t$ production, with
$t \to W^+b \to l \nu b $, $\bar t \to W^- \bar b \to jj \bar b$, when one of
the jets resulting from the $W^-$ decay is missed by the detector and the other 
is mistagged as a $b$ jet. Another small background is $W \bb j$ production.

The $Ht$ process has a smaller cross section than
$t \bar t$ with subsequent decay $\bar t \to H \bar q$ for the same
values of $g_{tq}$, but also receives an additional contribution from the
latter when $q$ is missed by the detector. This correction is not very
significant
for the analogous case of $Zt$ production,
about $+10\%$ for the up quark and $+50\%$ for
the charm, but in this case with $M_H > M_Z$ the additional contribution becomes
more important. In our calculations we will then take into account both
processes. The signals and backgrounds are generated as for the $\bbbj$ 
case. To estimate the number of events in which a jet is missed by the detector,
we consider that a jet is missed when it has a low transverse momentum $p_T^j < 
20$ GeV or a very large pseudorapidity $\eta^j > 3$.

The signal is reconstructed choosing first the pair of $b$ jets whose invariant
mass is closest to $M_H$. The remaining $b$ jet is assigned to the decay of the
top quark, whose mass is reconstructed as in the previous Section. To improve
the signal to background ratio we use the kinematical cuts shown in
Table~\ref{tab:3}.

\begin{table}[htb]
\begin{center}
\begin{tabular}{cc}
Variable & Cut \\
$\mhrec$ & 115--125 \\
$\mtrec$ & 160--190 \\
$H_T$ & $>220$ \\
\end{tabular}
\caption{Kinematical cuts for the $\bbb$ signal.
The masses and the energy are in GeV.
\label{tab:3}}
\end{center}
\end{table}

In this case we cannot reduce the background imposing a veto cut to reject
events in which a pair of jets has an invariant mass around $M_W$, because
one of these jets is missed by the detector. Using an ANN does not improve
the situation. The number of events for the $gu \to Ht$ and
$gc \to Ht$ signals, the additional contribution $t \bar t \to Ht(q)$
and their backgrounds is collected in Table~\ref{tab:4}.

\begin{table}[htb]
\begin{center}
\begin{tabular}{ccccc}
 & \multicolumn{2}{c}{Run L} & \multicolumn{2}{c}{Run H} \\
& before & after & before & after \\[-0.4cm]
& cuts & cuts & cuts & cuts \\
$gu \to Ht$ & 57.2 & 35.1 & 470 & 293 \\
$gc \to Ht$ & 11.9 & 6.8 & 95 & 56 \\
$t \bar t \to Ht(q)$ & 42.1 & 22.0 & 170 & 186 \\
$t \bar t$ & 1373 & 143 & 11260 & 1190 \\
$W \bb j$ & 77.2 & 1.6 & 619 & 12.7
\end{tabular} 
\caption{Number of $\bbb$ events before and after
the kinematical cuts in Table~\ref{tab:3} for the $Ht$ signal, the additional 
contribution $t \bar t \to Ht(q)$ and their backgrounds $t \bar t$ and
$W \bb j$. We use $g_{tq}=0.2$.
\label{tab:4}}
\end{center}
\end{table}

The limits obtained from the $Ht$ process are much less
restrictive than from top decays. If no signal is observed, we obtain
from Table~\ref{tab:4} $g_{tu} \leq 0.13$, $g_{tc} \leq 0.19$ at Run L and
$g_{tu} \leq 0.076$, $g_{tc} \leq 0.11$ at Run H. The same is true for Tevatron
Run II, where the bounds obtained are very poor, $g_{tu} \leq 0.40$,
$g_{tc} \leq 1.2$

\section{LHC discovery potential}
\label{sec:4}

In the previous Sections we have considered that the signal is not seen, and
have obtained 95\% CL upper bounds on the size of the FCN couplings. 
These bounds imply the limits in the branching ratios of  $t \to Hq$ in
Table~\ref{tab:5}.

\begin{table}[htb]
\begin{center}
\begin{tabular}{ccccccc}
& \multicolumn{2}{c}{LHC Run L} & \multicolumn{2}{c}{LHC Run H}
& \multicolumn{2}{c}{Tevatron Run II} \\
Signal & $\mathrm{Br}(t \to Hu)$ & $\mathrm{Br}(t \to Hc)$ &
$\mathrm{Br}(t \to Hu)$ & $\mathrm{Br}(t \to Hc)$ &
$\mathrm{Br}(t \to Hu)$ & $\mathrm{Br}(t \to Hc)$ \\
$\bbbj$ & $1.4 \times 10^{-4}$ & $1.4 \times 10^{-4}$
& $4.5 \times 10^{-5}$ & $4.5 \times 10^{-5}$
& $5.2 \times 10^{-3}$ & $5.2 \times 10^{-3}$ \\
$\bbb$ & $6.6 \times 10^{-4}$ & $1.3 \times 10^{-3}$
& $2.2 \times 10^{-4}$ & $4.3 \times 10^{-4}$ 
& $5.9 \times 10^{-3}$ & $5.3 \times 10^{-2}$\\
\end{tabular}
\caption{95\% CL upper limits on branching fractions of anomalous top decays
obtained from the two processes analyzed. We also include the results for
Tevatron Run II for comparison.
\label{tab:5}}
\end{center}
\end{table}

The limits obtained for different values of $M_H$ are very similar. For
a lighter Higgs, the larger phase space (in the case of top decays) or the 
larger structure functions (in $Ht$ production) increase the signal cross
section and hence allow a better determination of $g_{tq}$. This improvement
disappears when expressing the limit in terms of a branching ratio. After
repeating the analysis using $M_H=110$, $\mathrm{Br}(H \to \bb) = 0.8$, we find
limits $3\%$ smaller from top decays and $5\%$ smaller from $Ht$
production. A heavier Higgs has the opposite behaviour, with the disadvantage
of a smaller branching ratio $H \to \bb$. With $M_H=130$,
$\mathrm{Br}(H \to \bb)=0.45$ the limits from top decays and $Ht$ production
are 1.5 and 1.3 times larger, respectively.

Let us turn now to the possibility of observing these anomalous decays.
The most common criterion to estimate the discovery potential is to use
the discovery significance defined as the ratio of the expected signal
divided by the expected background fluctuation, $s \equiv n_s/\sqrt{n_b}$.
This is adequate when $n_b > 5$ and the Poisson distribution of the background
can be approximated by a Gaussian with standard deviation $\sqrt{n_b}$. One has
``evidence'' ($3 \, \sigma$) when  $s > 3$, and ``discovery''
($5 \, \sigma$) occurs with $s > 5$.
 
In a general 2HDM, the most common {\em ansatz} for the FCN couplings is to
assume that they scale with the quark masses. Under this assumption,
the expected size of these couplings is
$g_{tc} \simeq \sqrt{m_t m_c}/M_W = 0.20$,
$g_{tu} \simeq \sqrt{m_t m_u}/M_W = 0.012$. An $Htc$ coupling of this
size (leading to $\mathrm{Br}(t \to Hc)=1.5 \times 10^{-3}$)
will easily be detected
and precisely measured. In top decays, at Run H we have $s = 48$ using standard
cuts and $s = 68$ using the network cut. The benefit of the ANN is clear: the
luminosity required to discover a signal is reduced to one half.
In $Ht$ production, the significance is much smaller, $s = 7$. 
The minimum coupling which can be discovered with $5 \, \sigma$ is 
$g_{tc} = 0.054$ ($\mathrm{Br}(t \to Hc)=1.1 \times 10^{-4}$), and the minimum
coupling which can be seen with $3 \, \sigma$ is
$g_{tc} = 0.042$ ($\mathrm{Br}(t \to Hc)=6.5 \times 10^{-5}$). 
An $Htu$ coupling $g_{tu} = 0.012$ is too small to be seen, because in
top decays we have only $s = 0.23$, and in $Ht$ production $s = 0.046$. Note
that for a 2HDM this {\em ansatz} implies that there are other interesting
FCN current processes, for instance $H \to \mu \tau$ at a muon collider
\cite{papiro30}.

Another possibility is to consider that the FCN couplings between two quarks are
proportional to some CKM mixing angles involving these quarks. Rephasing
invariance implies that these couplings are of the form
$g_{qq'} = \lambda V_{qi} V_{q'i}^*$, with $i=d,s,b$ \cite{papiro3b,papiro4}.
Choosing $i=b$, the hierarchy among the
mixing angles guarantees a small value for $g_{uc} = \lambda V_{ub} V_{cb}^*$.
With $\lambda = 5$ we recover the values of $g_{uc}$ and $g_{tc}$ of the
previous {\em ansatz}, but $g_{tu}$ is a factor of two larger, $g_{tu} =
0.025$. An Htu coupling of this size is now near the detectable level: 
$s=1.1$ in top decays and in $s=0.22$ in $Ht$
production, and if it is enhanced by a factor $\sim 2$ it
would also be detected.

In models with exotic quarks the top FCN scalar couplings can be large, $g_{tu}
\simeq 0.31$ or $g_{tc} \simeq 0.35$ (but not both simultaneously). These are
the only models
where the top can naturally have a large mixing with the up quark. However, the 
effects of top mixing in these models are most clearly seen with the appearance
of tree level $Ztq$ couplings, leading to top anomalous decays $t \to Zq$
\cite{papiro27b} and $Zt$ production \cite{papiro27}. 
If these processes are not observed at LHC, the $Htq$ couplings
(proportional to the FCN $Ztq$ couplings) 
must be small, $g_{tu} \leq  0.024$, $g_{tc} \leq 0.030$,
below the detectable level.

To analyze the case of top decays $t \to Hq$ in the MSSM it is necessary to
recalculate the signal and backgrounds and train again the ANN with the new
data. The reason is that in the region of parameters of interest the Higgs
width is much larger due to the enhanced $Hbb$ coupling, and the $\mhrec$
distribution is broader in principle.
For consistency we will use the parameters of Ref. \cite{papiro5}:
$\tan \beta = 35$, a pseudoscalar mass $M_A = 120$ GeV and
$c_v = c_a = 1/\sqrt 2$. With these parameters, the Higgs mass is in the
interval $M_H = 105-120$ GeV \cite{papiro31}. The
kinematics is fairly independent of the particular value of $M_H$ used, as we
have shown above, the only difference being the Higgs decay rate to $b \bar b$
which varies
slightly in this mass range. For our analysis we take $M_H = 110$ GeV,
$\mathrm{Br}(H \to \bb) = 0.9$ 
\cite{papiro28}. Our limits on $\mathrm{Br}(t \to Hq)$
are proportional to $1/\mathrm{Br}(H \to \bb)$.
To separate signal from
background we select events with network output $r>0.994$.

There are large regions in the MSSM parameter space where $\mathrm{Br}(t \to
Hc)$ is greater than $1.1 \times 10^{-4}$ and even reaches $4 \times 10^{-4}$
\cite{papiro5}. These decays can be discovered at LHC Run H with
$5 \, \sigma-18 \, \sigma$. Also, a rate
$\mathrm{Br}(t \to Hc) = 6.7 \times 10^{-5}$
is observable with $3 \, \sigma$. The decays $t \to Hu$ have a too small rate
to be seen.

On the other hand, if no signal is
observed we obtain a limit $g_{tq} \leq 0.054$ at Run L and $g_{tq} \leq 0.031$
at Run H. Note that although these limits seem better than those obtained for
$M_H=120$, when translated into branching ratios they give very similar bounds,
$\mathrm{Br}(t \to Hq) \leq 1.4 \times 10^{-4}$ at Run L and
$\mathrm{Br}(t \to Hq) \leq 4.7 \times 10^{-5}$ at Run H.

\begin{ack}
We are indebted to J. Guasch for discussions and for providing us with data
for the MSSM Higgs analysis. We have also benefited from discussions with
F. del Aguila, Ll. Ametller and J. Seixas. J. A. A. S. thanks the members of the
CFIF for their hospitality during the realization of this work.
This work was partially supported by CICYT under contract AEN96--1672 and by the
Junta de Andaluc\'{\i}a, FQM101.
\end{ack}

\end{document}